\documentclass[aps,prb,reprint]{revtex4-2}
\usepackage{mathrsfs}
\usepackage{amsmath}
\usepackage{gensymb}
\usepackage{amsfonts}
\usepackage{amssymb}
\usepackage{amsthm}
\usepackage{graphicx}
\usepackage{natbib}
\usepackage{xcolor}
\usepackage{hyperref}
\usepackage{bm}
\usepackage[caption=false]{subfig}
\usepackage{verbatim}
\usepackage{siunitx}
\usepackage{tabularx}
\usepackage{makecell}
\begin{document}

\title{Barium Hexaferrite Thin Films as a Scalable Magnetic-Insulator\\ Platform for Proximity-Engineered Spintronics}
\author{Shyam Sundar Poriah$^1$,  Sanjana D. S.$^2$, Agrim Sharma$^3$, Sreelakshmi M. Nair$^4$, Pankaj Bhardwaj$^1$, Laxmipriya Nanda$^1$,  Aryaman Das$^1$, Jagadish Rajendran$^1$,   R. S. Patel$^4$, Manish Jain$^3$ and Dhavala Suri$^{1,}$}
\email{dsuri@iisc.ac.in}
\affiliation{$^1$Centre for Nanoscience and Engineering, Indian Institute of Science, Bengaluru, Karnataka 560012, India} 
\affiliation{$^2$Department of Materials Engineering, Indian Institute of Science, Bengaluru, Karnataka 560012, India}
\affiliation{$^3$Department of Physics, Indian Institute of Science, Bengaluru, Karnataka 560012, India} 
\affiliation{$^4$Department of Physics, Birla Institute of Technology \& Science Pilani - K K Birla Goa Campus, Zuarinagar, Goa 403726, India}

\begin{abstract}
Rare-earth iron garnets, such as yttrium iron garnet (YIG) and thulium iron garnet (TmIG), are the benchmark magnetic insulators for spintronic and magnonic devices, but achieving usable perpendicular magnetic anisotropy (PMA) in these materials typically relies on substrate strain-engineering, requiring careful lattice-matching and specific growth conditions that constrain material accessibility. Here we establish sputter grown barium hexaferrite (BaFe$_{12}$O$_{19}$, BaM) as a magnetic-insulator alternative with strong intrinsic perpendicular anisotropy, requiring no strain engineering. X-ray diffraction, transmission electron microscopy and Raman spectroscopy confirm stoichiometric films with atomically smooth surfaces, while first-principles calculations corroborate a robust ferrimagnetic ground state. The films exhibit square out-of-plane hysteresis with a coercive field of $\approx$~0.1~T. Unlike rare-earth garnets  \cite{Kubota2012}, the perpendicular anisotropy in BaM is intrinsic to its magnetoplumbite crystal structure, arising independent of highly ordered strain. Interfaced with Pt and with exfoliated BiSbTeSe$_{2}$ (BSTS), BaM induces proximity induced anomalous Hall transport, confirming efficient interfacial exchange coupling, while the BSTS/BaM heterostructure shows an additional Hall contribution suggestive of non-collinear interfacial spin textures. These results position BaM thin films as a scalable magnetic-insulator platform for spintronic and topological heterostructure devices beyond the constraints of garnet chemistry.
\end{abstract}

\maketitle

Perpendicular magnetic anisotropy (PMA) is essential for high-density magnetic storage, spin--orbit torque devices, and nanoscale spintronic applications\cite{Dieny2017,Ikeda2010,Johnson1996}. Conventional metallic multilayers such as Co/Pt\cite{Nakajima1998}, Co/Pd\cite{Agui2004,Johnson1996-hl,denBroeder1991}, and CoFeB\cite{Ikeda2010} achieve PMA through interfacial spin--orbit coupling, but intrinsic electronic dissipation in metallic ferromagnets limits spin diffusion length and energy efficiency. Magnetic insulators (MIs) circumvent this limitation, enabling magnetization dynamics, spin pumping, and spin--orbit-torque-driven switching free of parasitic charge-current losses\cite{Avci2016,Vlez2019,Wu2018,Avci2019}. Garnet ferrites such as yttrium iron garnet (YIG)\cite{Serga2010,Schmidt2020,Askarzadeh2025} and thulium iron garnet (TmIG)\cite{Ciubotariu2019,Tang2016} have been extensively studied for their low magnetic damping, yet they require rare-earth elements, possess complex crystallographic structures, and typically depend on substrate-induced strain to realize usable PMA, constraining material accessibility and growth scalability.

Barium hexaferrite (BaFe$_{12}$O$_{19}$, BaM), a magnetoplumbite-structured ferrimagnetic insulator, offers a compelling alternative: strong intrinsic uniaxial anisotropy along the crystallographic $c$-axis\cite{Krichevtsov2023}, a high Curie temperature ($\sim$740~K)\cite{Shirk1969}, high electrical resistivity ($\sim$10$^{8}$~$\Omega\cdot$cm)\cite{Marouani2021}, and large magnetocrystalline anisotropy arising from five inequivalent Fe$^{3+}$ sublattices\cite{Pullar2012}. These properties, combined with its chemical stability and cost-effective synthesis, have long made BaM attractive for permanent magnets, microwave absorbers\cite{Qiu2005}, and magneto-optical recording\cite{Masterson1993}, and more recently as a magnetic-proximity source in heavy-metal\cite{Li2016} and topological-insulator\cite{Li2020} heterostructures. However, realizing phase-pure, highly $c$-axis-oriented BaM thin films remains challenging: the multi-sub-lattice ferrite structure demands precise control of stoichiometry, oxygen environment, and crystallization kinetics. Further, secondary iron-oxide phases readily suppress anisotropy and degrade interfacial quality.

\begin{figure*}[t]
\centering
\includegraphics[width=\linewidth]{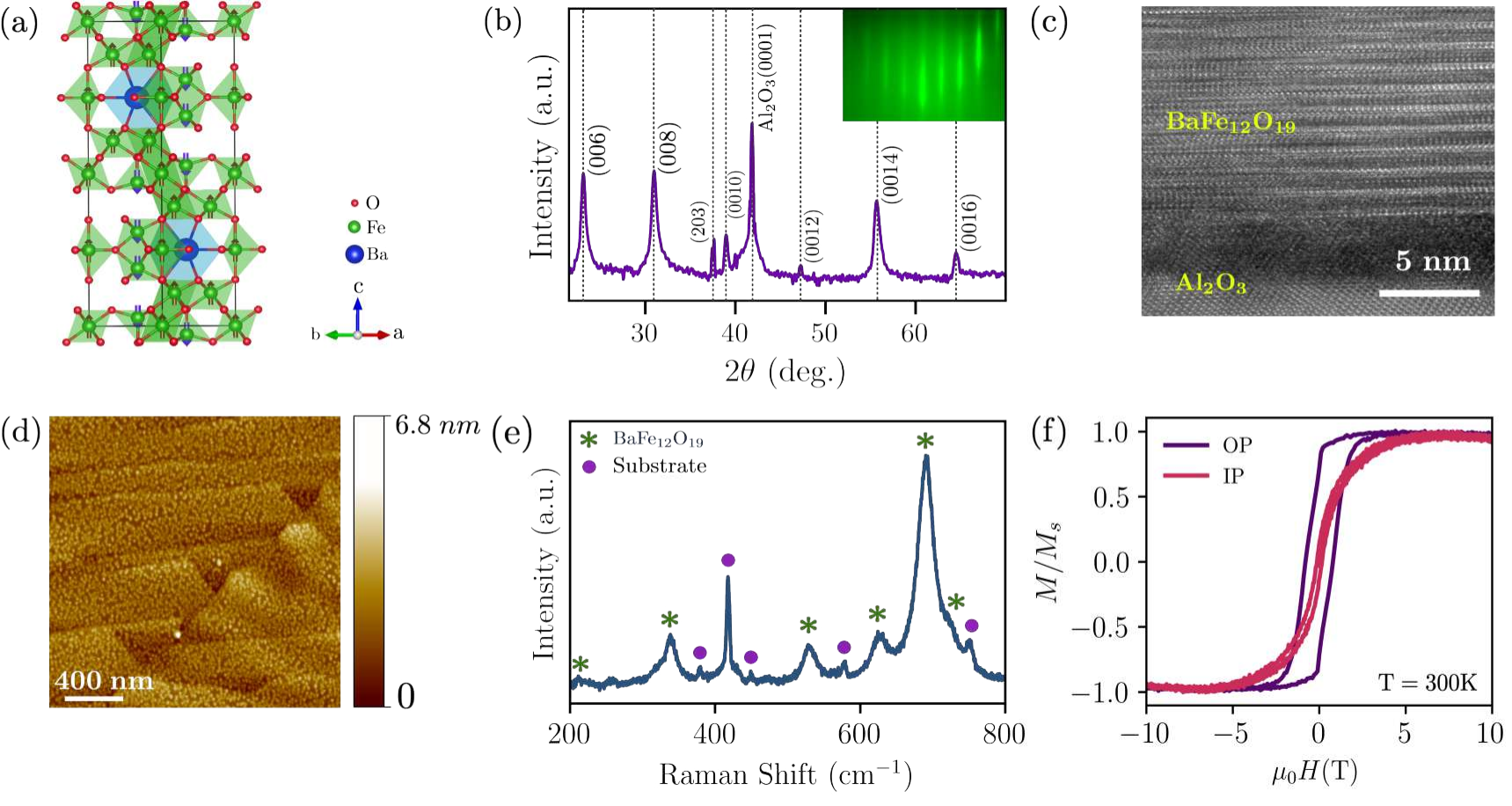}
\caption{(a) Crystal structure of BaM generated using \textit{VESTA}. (b) XRD spectrum of BaM thin film(BFO53); the inset shows the RHEED pattern, exhibiting streaky features indicative of highly ordered growth. (c) Cross-sectional TEM image of the BaM film on Al$_2$O$_3$(0001), showing ordered highly ordered growth along the $c$-axis. (d) AFM image of the BaM film surface, displaying smooth, terrace-like features. (e) Raman spectrum of the film, with peaks assigned to the film and substrate indicated in the legend. (f) Magnetization as a function of magnetic field measured out-of-plane (OP) and in-plane (IP) at room temperature. }
\label{fig:1}
\end{figure*}

BaM thin films have been grown by sol--gel processing\cite{Pullar2002,Chawla2014}, pulsed laser deposition\cite{Zhou2021,Carosella1992,Lisfi2003,Liu2004}, liquid-phase epitaxy\cite{Wang2002,Yoon2003,Kranov2006}, laser MBE\cite{Krichevtsov2023,Krichevtsov2024}, ion-beam deposition\cite{Kostishin2021,VG2022}, and magnetron sputtering\cite{Zhang2019,Zhang2010,Xu2013}. While PLD has achieved epitaxial films with near-bulk magnetic properties, sputtering offers superior scalability, uniformity, and compatibility with industrial thin-film processing, yet phase-pure sputtered BaM with strong PMA and device-quality interfaces remains comparatively unexplored. High crystalline quality alone, however, does not establish spintronic relevance: BaM must also drive robust exchange-mediated proximity effects in adjacent non-magnetic layers. To demonstrate this, we engineer two representative heterostructures --- a Pt thin film and an exfoliated BiSbTeSe$_2$ (BSTS) topological-insulator flake, both interfaced with sputtered BaM --- providing a stringent test of BaM's efficacy as a proximity source for strongly spin--orbit-coupled materials.

BaFe$_{12}$O$_{19}$ thin films were deposited at $600\,^{\circ}\mathrm{C}$ by RF magnetron sputtering at a base pressure of $\approx 10^{-7}$ mbar and a working pressure of $5 \times 10^{-3}$ mbar. Following deposition, the films were annealed in air at $1000\,^{\circ}\mathrm{C}$ for 1 hour with a heating and cooling rate of 2K/min. 
Fig.~\ref{fig:1}(a) shows the hexagonal BaM structure modeled using the \textit{VESTA} software. The unit cell consists of alternating spinel-type ($S$) and hexagonal ($R$) blocks stacked along the $c$-axis: three $R$ blocks, each comprising two O$_4$ layers sandwiching a BaO$_3$ layer with composition Ba$^{2+}$Fe$^{3+}_6$O$^{2-}_{11}$, and two $S$ blocks, each comprising two O$_4$ layers with composition Fe$^{3+}_6$O$^{2-}_8$; successive blocks of each type are related by a 180$^\circ$ rotation about the $c$-axis\cite{Moitra2014}. Fig.~\ref{fig:1}(b) shows X-ray diffraction (XRD) $\theta$--$2\theta$ patterns of BaM thin films grown on $c$-cut Al$_2$O$_3$(0001) substrates under deposition conditions ranging from mixed Ar/O$_2$ atmospheres to pure Ar.  All diffraction peaks index to single-phase hexagonal  BaM with lattice constants $a$~=~\SI{5.813}{\angstrom} and $c$~=~\SI{23.076}{\angstrom}, with no evidence of secondary Fe- or Ba-based impurity phases, confirming the structural integrity of the films. The sharp diffraction peaks and low FWHM values further attest to the excellent crystalline quality of the films (see Supplementary Information for details). Fig.~\ref{fig:1}(c) shows cross-sectional transmission electron microscopy (TEM) confirming highly ordered growth of the BaM thin film on the sapphire substrate (inset Fig. 1 (b) shows the corresponding reflection high-energy electron diffraction pattern).

\begin{figure}[!ht]
\centering
\includegraphics[width=8cm]{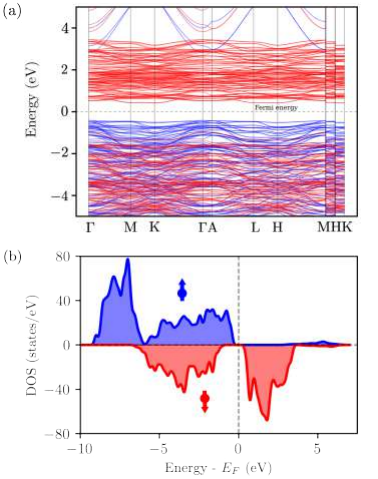}
\caption{(a) Spin-polarized electronic band structure and (b) density of states computed using density functional theory (DFT), with blue and red curves denoting the spin-up and spin-down channels, respectively.}
\label{fig:2}
\end{figure}

Fig.~\ref{fig:1}(d) presents the atomic force microscopy (AFM) image of the grown BaM thin films, exhibiting a root-mean-square roughness ($R_q$) of $\approx$~0.643~nm over a 2~$\times$~2~$\mu$m$^2$ region. The AFM image reveals a highly ordered thin-film growth morphology\cite{Nergis2023}, with a smooth, homogeneous surface exhibiting terrace-like features that mirror the atomic terraces of the underlying substrate. Isolated triangular protrusions are observed superimposed on this terraced background, consistent with preferentially crystallized orientations nucleating during growth and subsequently stabilized by thermally driven reordering during post-growth annealing\cite{Kishor2024}. The low surface roughness confirms high crystalline quality and smooth film growth, consistent with the structural results in Fig.~\ref{fig:1}(b,c). Raman spectroscopy further confirms the hexagonal BaM phase [Fig.~\ref{fig:1}(e)], evidenced by the Raman-active $E_{1g}$, $E_{2g}$, and $A_{1g}$ phonon modes\cite{Manglam_2023}. The absence of any secondary phases or oxide impurities in the Raman spectrum confirms the growth of high-quality BaM thin films.

\begin{figure*}[t]
    \centering
    \includegraphics[width=\linewidth]{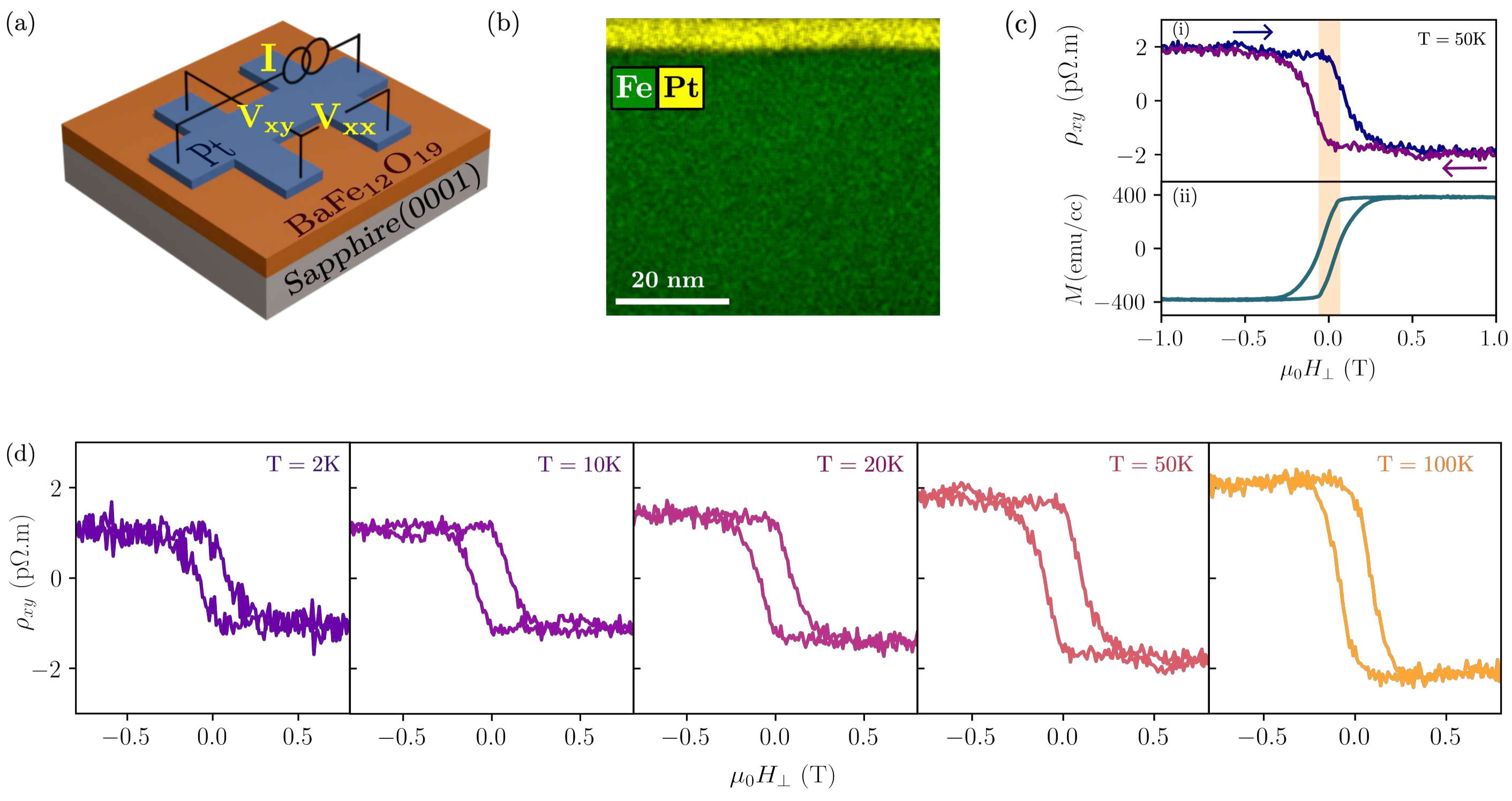}
    \caption{(a) Schematic of the Hall-bar device geometry used for the measurements. (b) Energy-dispersive X-ray spectroscopy (EDS) map of the Pt/BaM interface, showing no inter-diffusion of Fe into Pt. The dark yellow spots in Pt are due to background. (c)(i) Anomalous Hall resistivity of Pt(5~nm)/BaM as a function of magnetic field, and (ii) out-of-plane $M(H)$ of BaM. (d) Anomalous Hall resistivity at different temperatures, as indicated in the legend.}
    \label{fig:3}
\end{figure*}

Figs.~2(a) and 2(b) show the spin-polarized electronic band structure and spin-polarized density of states, respectively, calculated from first-principles DFT using the PBE+U method with a Hubbard $U = 3.7$~eV applied to the Fe $3d$ orbitals \cite{liyanage_theory_2013, cococcioni_linear_2005, timrov_pulay_2020}. The calculated band structure and density of states reveal a strongly spin-split ferrimagnetic ground state, with majority- and minority-spin states clearly separated near the Fermi level. These results support the large intrinsic magnetization of BaM and provide a microscopic basis for the strong perpendicular magnetic anisotropy observed in the films\cite{Moitra2014}. Calculations were performed using Quantum ESPRESSO\cite{giannozzi_quantum_2009, giannozzi_advanced_2017} with a $5\times5\times1$ shifted Monkhorst--Pack $k$-point grid, a plane-wave basis cutoff of 110~Ry for the wavefunctions, and Optimized Norm-Conserving Vanderbilt (ONCV) pseudopotentials from the PseudoDojo library\cite{van_setten_pseudodojo:_2018, hamann_optimized_2013}. As noted above, the magnetic structure of BaM arises from five crystallographically distinct Fe$^{3+}$ sublattices, distributed among three octahedral sites ($2\text{a}$, $4\text{f}_2$, $12\text{k}$), one tetrahedral site ($4\text{f}_1$), and one trigonal bipyramidal site ($2\text{b}$). Spin-polarized calculations were performed for several candidate spin configurations to determine the ground-state magnetic structure. In the ground state, the magnetic moments on the $4\text{f}_1$ and $4\text{f}_2$ Fe sites are antiparallel to those on the $2\text{a}$, $12\text{k}$, and $2\text{b}$ Fe sites, yielding a net magnetization of $40\,\mu_\text{B}$ per unit cell.

The magnetization $M$ of the thin film grown was measured as a function of the external magnetic field $H$ using a vibrating sample magnetometer in the range of -1~T to 1~T, showing in-plane (IP) and out-of-plane (OOP) hysteresis, as shown in Fig.\ref{fig:1} (f). A clear square-like hysteresis appears in the configuration where magnetic field is perpendicular to sample plane, where as the in-plane magnetic field configuration shows a non-saturating response with weak-hysteresis. This indicates that c-axis is the easy axis confirming the perpendicular magnetic anisotropy of the sample. We next investigate the most significant application of these ultra-smooth BaM films: proximity-induced interfacial coupling with materials possessing strong spin–orbit interaction. 

Heterostructures comprising BaM/Pt [BaM (55~nm)/Pt(5~nm)] were fabricated, with Pt chosen for its large spin--orbit coupling strength. Devices were patterned into Hall-bar geometries using focused ion beam (FIB) milling [Fig.~\ref{fig:3}(a)], avoiding conventional lithographic processing steps that could introduce chemical contamination or damage the pristine film surface. Energy-dispersive X-ray spectroscopy (EDS) mapping of the heterostructure [Fig.~\ref{fig:3}(b)] reveals a sharp elemental demarcation between Pt and Fe, indicating negligible interdiffusion of Fe into the Pt layer and thereby ruling out any spurious magnetic signature originating from Fe intermixing. The Pt film was deposited and maintained at room temperature or below throughout growth to further suppress interfacial intermixing.

Remarkably, Hall measurements reveal a clear anomalous Hall effect (AHE) in the BaM/Pt heterostructure [Fig.~\ref{fig:3}(c)], despite Pt alone exhibiting no intrinsic ferromagnetic order. This observation provides strong evidence for proximity-induced magnetic polarization of the interfacial electronic states through exchange coupling with the underlying BaM layer. The microscopic origin of spin transport and the anomalous Hall response in ferromagnetic-insulator (FMI)/Pt heterostructures maybe attributed to a few possibilities~\cite{Huang2012,Lu2013,Miao2014,Chen2013,Meyer2015}. Two principal mechanisms have been proposed: (i) a magnetic proximity effect (MPE) arising from interfacial Fe-$3d$--Pt-$5d$ orbital hybridization~\cite{Geprgs2012,Lu2013}, and (ii) a spin Hall effect--driven anomalous Hall contribution (SHE-AHE) governed by the interfacial spin-mixing conductance~\cite{Brataas2000,Tserkovnyak2002,Chen2013,Chen2016}. These two mechanisms exhibit contrasting temperature dependences: the MPE-induced Hall signal typically strengthens upon cooling, whereas the SHE-AHE contribution weakens with decreasing temperature, and their competition can produce a sign reversal of $R_{\mathrm{AHE}}$~\cite{Liang2018,Ding2021}, though the microscopic origin of such reversals remains debated~\cite{Nakayama2013,Weiler2013}. In addition, thermally excited magnons have been shown to mediate spin transport across FMI layers over micrometer length scales~\cite{Cornelissen2015,Goennenwein2015}, and the interplay between magnon-mediated and electronic spin-transfer processes at the FMI/Pt interface remains incompletely understood. While SHE-AHE and its temperature dependence have been extensively studied in YIG/Pt and TIG/Pt heterostructures~\cite{Nakayama2013,Weiler2013,Meyer2015}, the relative roles of SHE-AHE, MPE, and magnon-mediated transport remain largely unexplored in BaM-based heterostructures.

In the present BaM/Pt heterostructure, the anomalous Hall resistivity $\rho_{\mathrm{AHE}}$ increases monotonically with increasing temperature, with no sign reversal observed over the entire measured range [Fig.~\ref{fig:3}(d)]. This behavior is opposite to that expected for an MPE-dominated response, which typically strengthens upon cooling, and is instead consistent with an SHE-mediated origin, in which the spin current generated via the SHE in Pt is modulated by exchange coupling with the BaM interfacial magnetization through the spin-mixing conductance. The increasing thermal population of magnons at elevated temperatures may further enhance spin-angular-momentum transfer across the BaM/Pt interface, amplifying the SHE-mediated Hall response through more efficient interfacial spin transmission; although the present measurements do not directly isolate the magnon contribution, the observed temperature dependence is compatible with magnon-assisted interfacial spin transport. While a residual MPE contribution cannot be entirely excluded, the absence of a sign reversal and the monotonic strengthening with temperature together indicate that the SHE-AHE response dominates in this system. Magnetization measurements [Fig.~\ref{fig:3}(c)(ii)]  provide further support, with the observed coercive field consistent with the anomalous Hall response.

\begin{figure}[!ht]
\centering
\includegraphics[width=8 cm]{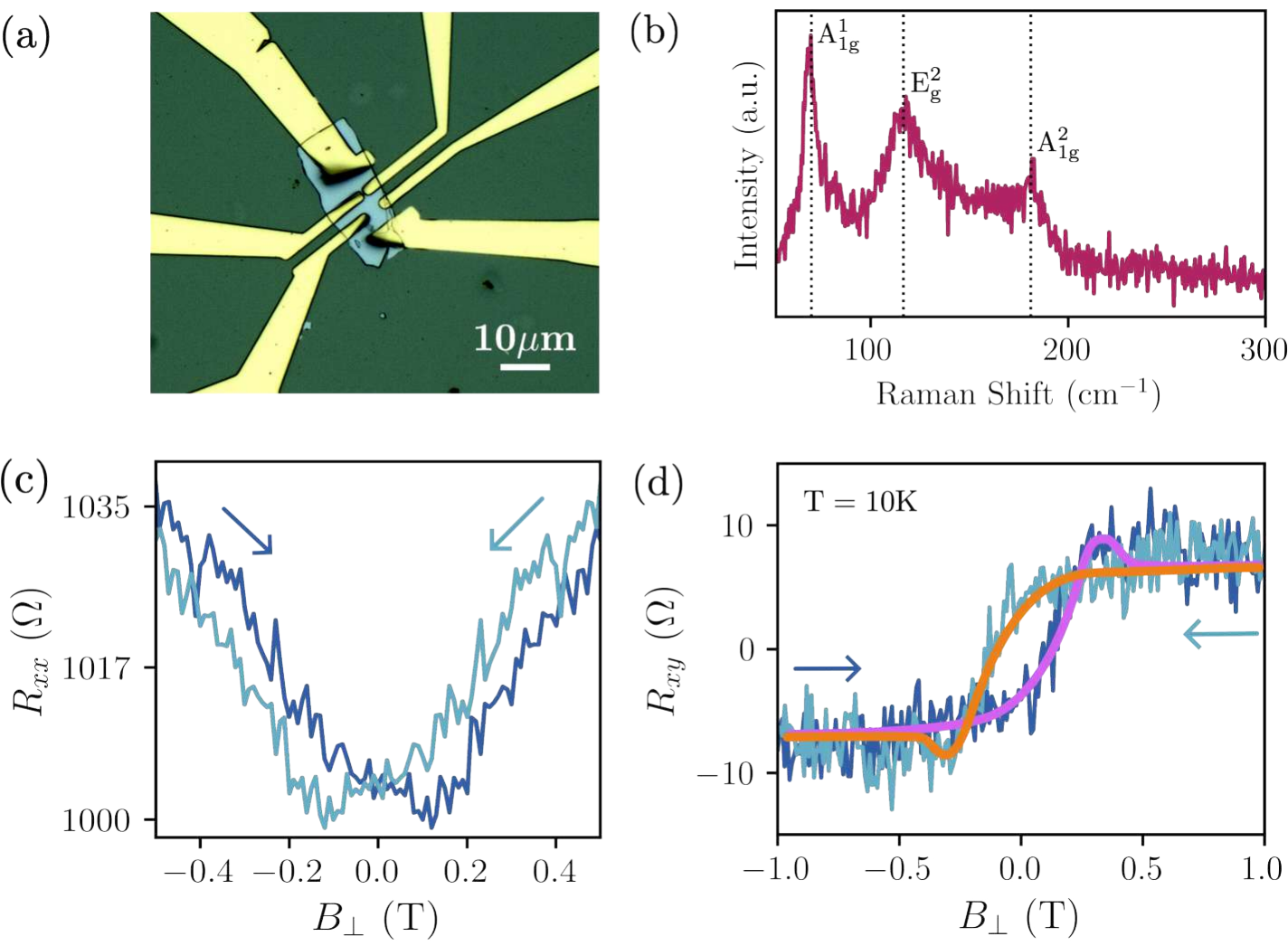}
\caption{(a) Optical image of the device fabricated on a BSTS flake transferred onto the BaM film. (b) Raman spectrum of the BSTS flake on BaM. Dotted lines indicate the  phonon modes corresponding to vibrations as indicated in the figure.   (c) $R_{xx}$ and (d) $R_{xy}$ as a function of $B_\perp$, measured in forward and reverse field sweep directions, as indicated by the arrows.  }
\label{fig:4}
\end{figure}

Finally, we investigate magnetic proximity coupling in an exfoliated flake of BSTS, a three-dimensional topological insulator with strong intrinsic spin--orbit coupling (SOC)\cite{Ren2011}. This heterostructure validates whether the interfacial exchange interaction induced by BaM can be observed in an exfoliated van der Waals topological material, while preserving the large SOC of its topological surface states. Fig.~\ref{fig:4}(a) shows an optical image of the BSTS device fabricated on BaM, with Ti/Au electrodes defined by standard electron-beam lithography; the BSTS flake composition is confirmed via Raman spectroscopy [Fig.~\ref{fig:4}(b)]. The Raman peaks of BaM do not manifest in this spectrum due to larger thickness (200~nm) of BSTS flakes.  At $T \approx 10$~K, the device exhibits pronounced hysteresis in the magnetoresistance [Fig.~\ref{fig:4} (c)] together with a clear AHE [Fig.~\ref{fig:4} (d)], demonstrating interfacial ferromagnetism through magnetic proximity coupling. Notably, the Hall response also contains a weak additional Hall contribution, manifesting as a hump-like feature superimposed on the dominant AHE background. We note that such humps in $\rho_{xy}(H)$ can, in principle, also arise from a superposition of two AHE channels \cite{Ou2025} with distinct coercive fields --- for instance, associated with magnetically inhomogeneous regions of the BSTS/BaM interface switching at different field values --- rather than from a genuine topological Hall effect (THE). While the present dataset does not allow us to unambiguously distinguish between these two scenarios, we discuss the feature here as a possible signature of non-collinear spin textures at the interface, motivating further investigation. This additional Hall component is instead consistent with interfacial symmetry breaking combined with the strong SOC of BSTS, which can stabilize canted spin configurations and non-collinear spin textures at the magnetic interface~\cite{Li2020}. Although the magnitude of the THE is modest, its presence suggests the formation of interfacial spin textures beyond a purely collinear ferromagnetic state.

In conclusion, these results decouple perpendicular magnetic anisotropy from the strain-engineered epitaxy that has constrained magnetic-insulator spintronics to garnet chemistry. Because BaM's anisotropy is a property of its crystal symmetry rather than an artifact of substrate matching, it survives the transition to a scalable growth technique and transfers its exchange coupling to both conventional and topological/spin-orbit coupled systems. This positions magnetoplumbite ferrites as a largely unexplored materials class for proximity-engineered spintronics, opening a route toward wafer-compatible magnetic-insulator platforms beyond the rare-earth constraint.

Authors are thankful to the National Nano-Fabrication Facility (NNFC) and the Micro and Nano Characterization Facility (MNCF) at the Centre for Nanoscience and Engineering, IISc. PB thanks Anusandhan National Research Foundation (ANRF), National Postdoctoral fellowship (PDF/2023/000444) for financial support.   DS thanks IISc start-up grant, Ministry of Electronics and Technology, Indian Space Research Organization, Infosys Foundation and Wadhwani Innovation Network for funding. Authors duly acknowledge funding from INOXCVA and INOX Airproducts for funding via CSR grants. RSP acknowledges funding from Department of Science and Technology, Govt. of India  for DST-FIST grant no. SR/FST/PS-I/2017/21.

\bibliography{abbreviation.bib,references.bib}

\clearpage

\setcounter{figure}{0}
\renewcommand{\thefigure}{S\arabic{figure}}

\setcounter{table}{0}
\renewcommand{\thetable}{S\arabic{table}}

\setcounter{equation}{0}
\renewcommand{\theequation}{S\arabic{equation}}

\begin{center}
    {\bf\Large SUPPLEMENTARY MATERIAL}
\end{center}

\section{Details about Barium Hexaferrite growth }

BaFe$_{12}$O$_{19}$ (BaM) thin films were grown by radio-frequency (RF) magnetron sputtering using a 99.99\% pure BaFe$_{12}$O$_{19}$ ceramic target. Prior to deposition, the sputtering chamber was evacuated to a base pressure of $2\times10^{-7}$ mbar. Single-crystalline Al$_2$O$_3$(0001) substrates were heated to 600 $^\circ$C and maintained at this temperature for 30 min to ensure thermal stabilization. The target was subsequently pre-sputtered for 15 min to remove surface contaminants and establish stable deposition conditions. Film growth was carried out at an RF power of 75 W under various sputtering conditions. To investigate the influence of the deposition atmosphere on the structural and magnetic properties of the films, the oxygen partial pressure was systematically varied by employing either mixed Ar/O$_2$ or pure Ar sputtering atmospheres while maintaining the deposition pressure accordingly. This optimization revealed that the sputtering environment has a pronounced effect on the crystalline quality and magnetic properties of the BaM thin films. Following deposition, all samples were post-annealed in ambient air at 1000 $^\circ$C for 1 h to promote crystallization of the magnetoplumbite phase. The temperature was increased from 600 $^\circ$C to 1000 $^\circ$C at a ramp rate of 2 K/min and subsequently cooled back to 600 $^\circ$C at the same rate. The deposition conditions and film thicknesses of all samples are summarized in Table~\ref{tab:deposition}.

\begin{table}[htbp]
\centering
\renewcommand{\arraystretch}{1.3}
\setlength{\tabcolsep}{4pt}

\resizebox{\columnwidth}{!}{%
\begin{tabular}{|c|c|c|c|c|}
\hline
\textbf{Sample No.} &
\makecell{\textbf{Deposition}\\\textbf{Pressure}\\\textbf{(mbar)}} &
\makecell{\textbf{Ar flow}\\\textbf{(sccm)}} &
\makecell{\textbf{O$_2$ flow}\\\textbf{(sccm)}} &
\makecell{\textbf{Thickness}\\\textbf{(nm)}} \\
\hline
BFO59 & $3\times10^{-2}$ & 30 & 3 & 36 \\
\hline
BFO55 & $3\times10^{-2}$ & 33 & -- & 34 \\
\hline
BFO53 & $5\times10^{-3}$ & 33 & -- & 55 \\
\hline
BFO61 & $5\times10^{-3}$ & 33 & -- & 165 \\
\hline
\end{tabular}%
}

\caption{Deposition parameters and film thickness of the samples.}
\label{tab:deposition}
\end{table}

Following deposition, all samples were post-annealed in ambient air at 1000 $^\circ$C for 1 h to promote crystallization of the magnetoplumbite phase. The temperature was increased from 600 $^\circ$C to 1000 $^\circ$C at a ramp rate of 2 K/min and subsequently cooled back to 600 $^\circ$C at the same rate.

\section{XRD analysis of different samples}

\begin{figure}[!ht]
\includegraphics[width=\columnwidth]{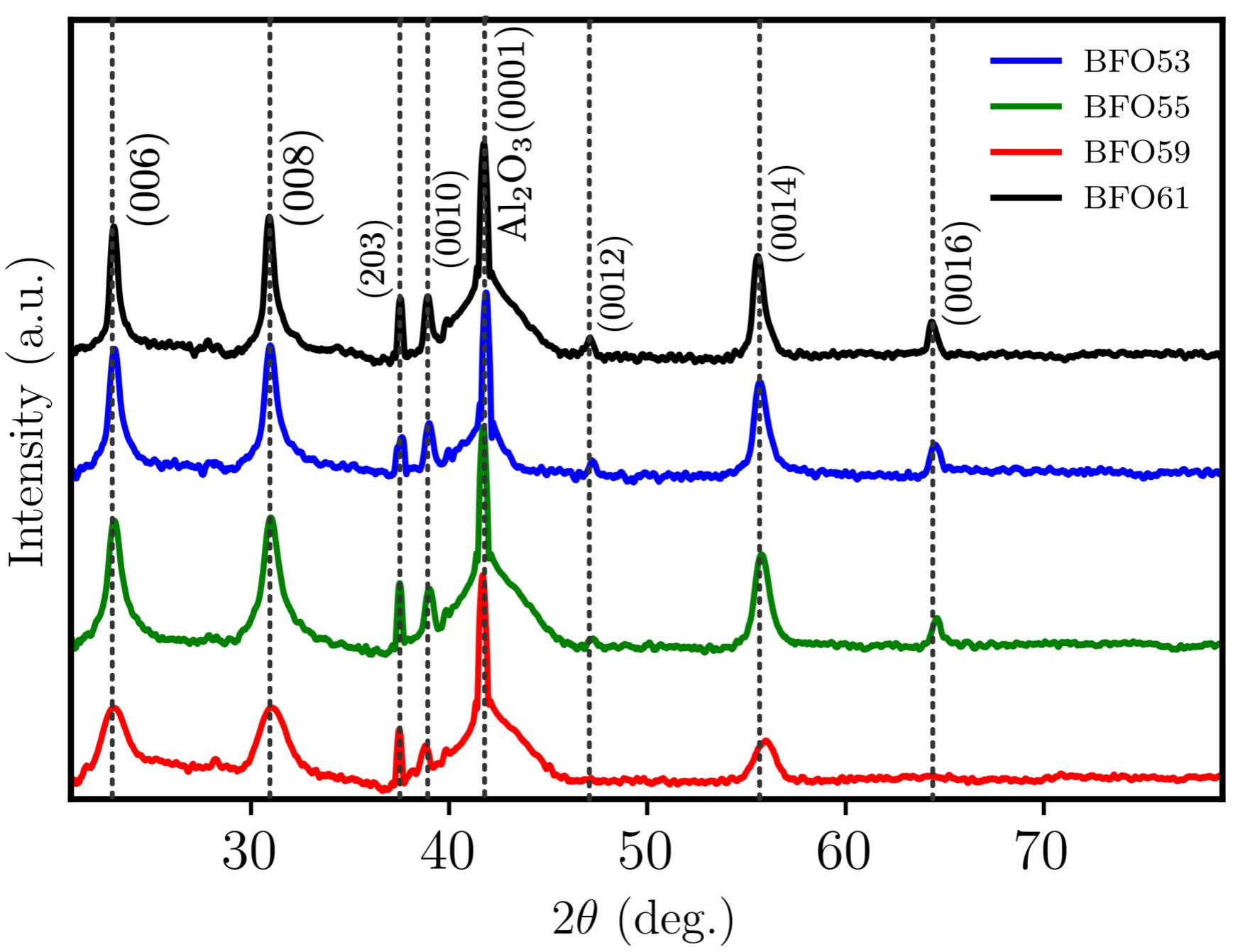}
\caption{X-ray diffraction (XRD) patterns of the BaFe$_{12}$O$_{19}$ thin films deposited under the growth conditions summarized in Table~\ref{tab:deposition}.}
\label{fig:S1}
\end{figure}

\begin{table}[htbp]
\centering
\renewcommand{\arraystretch}{1.4}

\resizebox{\columnwidth}{!}{%
\begin{tabular}{|c|c|c|c|}
\hline
\textbf{Sample No.} &
\makecell{\textbf{Lattice constant $a$}\\(\AA)} &
\makecell{\textbf{Lattice constant $c$}\\(\AA)} &
\makecell{\textbf{$\Delta c$ relative}\\\textbf{to bulk (\%)}} \\
\hline
Bulk BaFe$_{12}$O$_{19}$ & 5.89 & 23.18 & 0 \\
\hline
BFO59 & 5.828 & 23.016 & 0.72 \\
\hline
BFO55 & 5.824 & 23.057 & 0.54 \\
\hline
BFO53 & 5.813 & 23.076 & 0.46 \\
\hline
BFO61 & 5.818 & 23.114 & 0.29 \\
\hline
\end{tabular}%
}

\caption{Lattice parameters of BaFe$_{12}$O$_{19}$ thin films extracted from X-ray diffraction measurements. The percentage change in the out-of-plane lattice parameter $c$ is calculated with respect to the bulk value.}
\label{tab:lattice}
\end{table}

\begin{figure}[h!]
\includegraphics[width=\columnwidth]{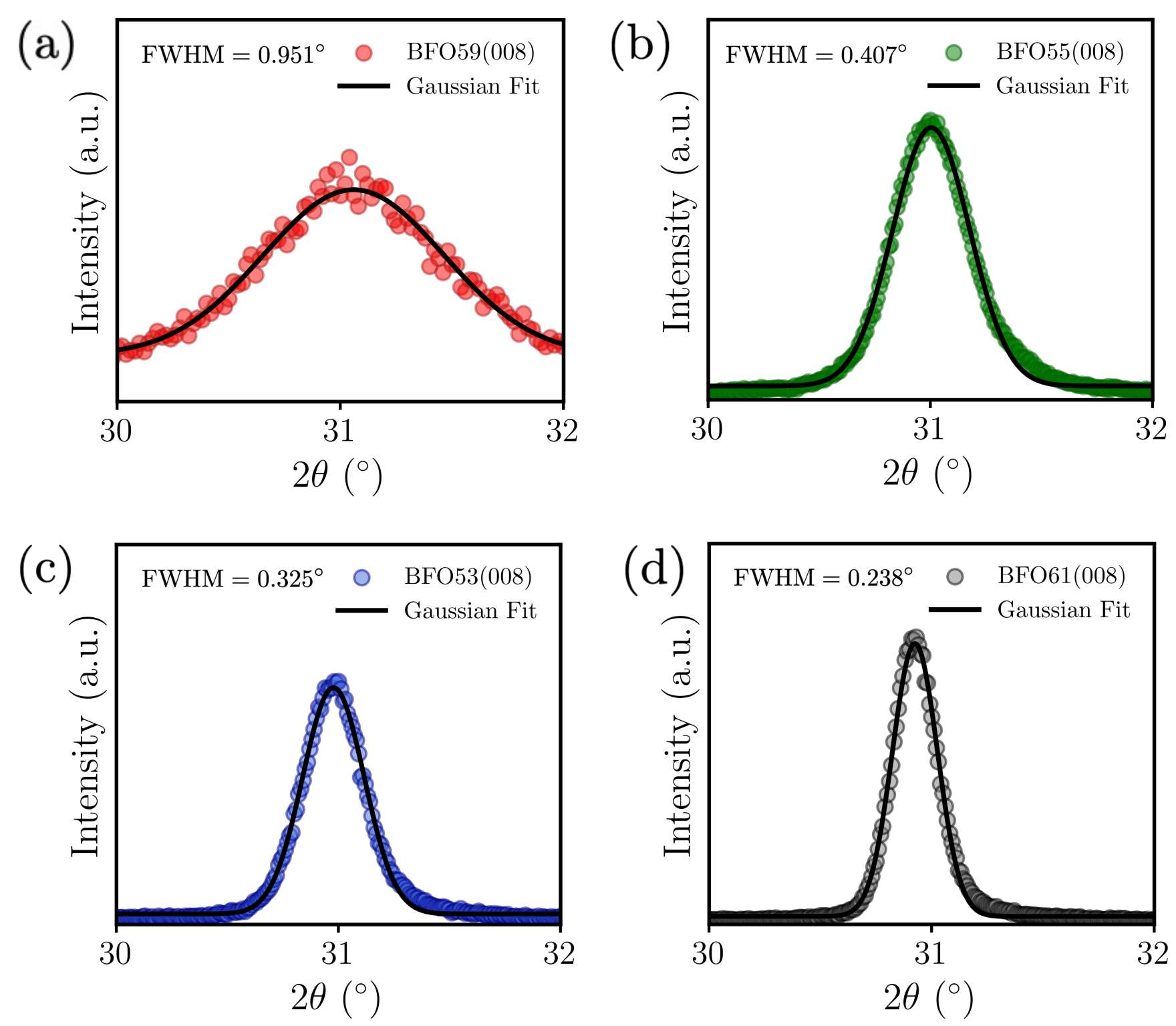}
\caption{Experimental and fitted X-ray diffraction profiles of the (008) peak for BaFe$_{12}$O$_{19}$ thin films. The full width at half maximum (FWHM) values obtained from the fits are shown in the figure.}
\label{fig:S2}
\end{figure}

The evolution of the lattice parameters indicates that all BaFe$_{12}$O$_{19}$ thin films possess a reduced out-of-plane lattice constant compared to the bulk value shown in Table 2, suggesting the presence of compressive strain. Among the investigated samples, BFO59 exhibits the largest contraction of the c-axis $\Delta c$=0.72$\%$ and a noticeably broadened (008) diffraction peak. The increased full width at half maximum (FWHM) points to reduced crystalline quality and may be associated with the presence of amorphous regions or secondary phases, such as $\alpha$-Fe$_{2}$O$_{3}$, promoted by the higher oxygen partial pressure during growth. In contrast, films deposited in a pure Ar atmosphere display significantly improved structural quality. Specifically, BFO55 and BFO53 show a gradual increase in the out-of-plane lattice parameter to 23.057 and 23.076 Å, corresponding to $\Delta c$ values of 0.54$\%$ and 0.46$\%$, respectively. This evolution is accompanied by a progressive narrowing of the (008) diffraction peak, indicating enhanced crystallinity with decreasing deposition pressure as shown in Figure 1. The thickest film, BFO61 (165 nm), exhibits lattice parameters closest to the bulk values, suggesting partial strain relaxation and the stabilization of phase-pure BaFe$_{12}$O$_{19}$. These results demonstrate that deposition in a pure Ar environment, combined with lower deposition pressure and increased film thickness, promotes improved crystallinity and structural properties in sputtered BaFe$_{12}$O$_{19}$ thin films.

\section{Topography(AFM) of different samples}

Among the investigated samples, BFO59 exhibits the lowest rms surface roughness; however, x-ray diffraction measurements indicate degraded crystalline quality and reduced phase purity, likely due to the increased oxygen partial pressure during deposition. In contrast, the thickest film, BFO61, displays excellent crystalline quality and lattice parameters close to the bulk values, but its relatively high surface roughness may limit proximity-induced interfacial effects. Samples BFO55 and BFO53 provide a favorable balance between structural quality and surface morphology, with rms roughness values below 1 nm and improved crystallinity. In particular, BFO53 combines excellent crystalline quality with the lowest surface roughness among the films, making it the most suitable candidate for heterostructure fabrication and proximity-induced spin transport studies.

\begin{figure}[!ht]
\includegraphics[width=\columnwidth]{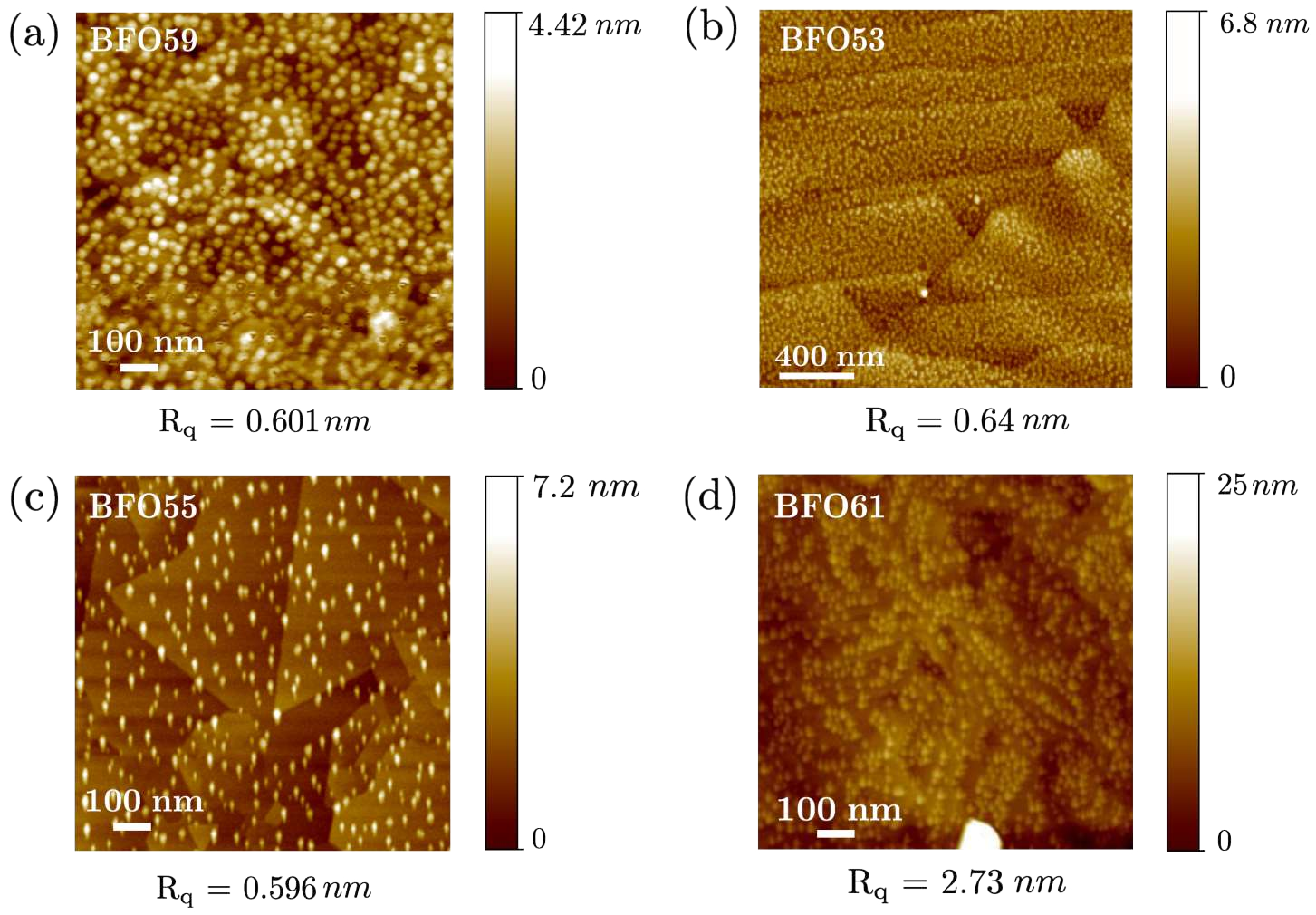}
\caption{Atomic force microscopy (AFM) images of BaFe$_{12}$O$_{19}$ thin films of different samples (sample ID indicated in the image). The corresponding root-mean-square(rms) surface roughness values are indicated in the figure.}
\label{fig:S3}
\end{figure}

\section{Raman analysis of different samples}

\begin{figure}[h!]
\includegraphics[width=\columnwidth]{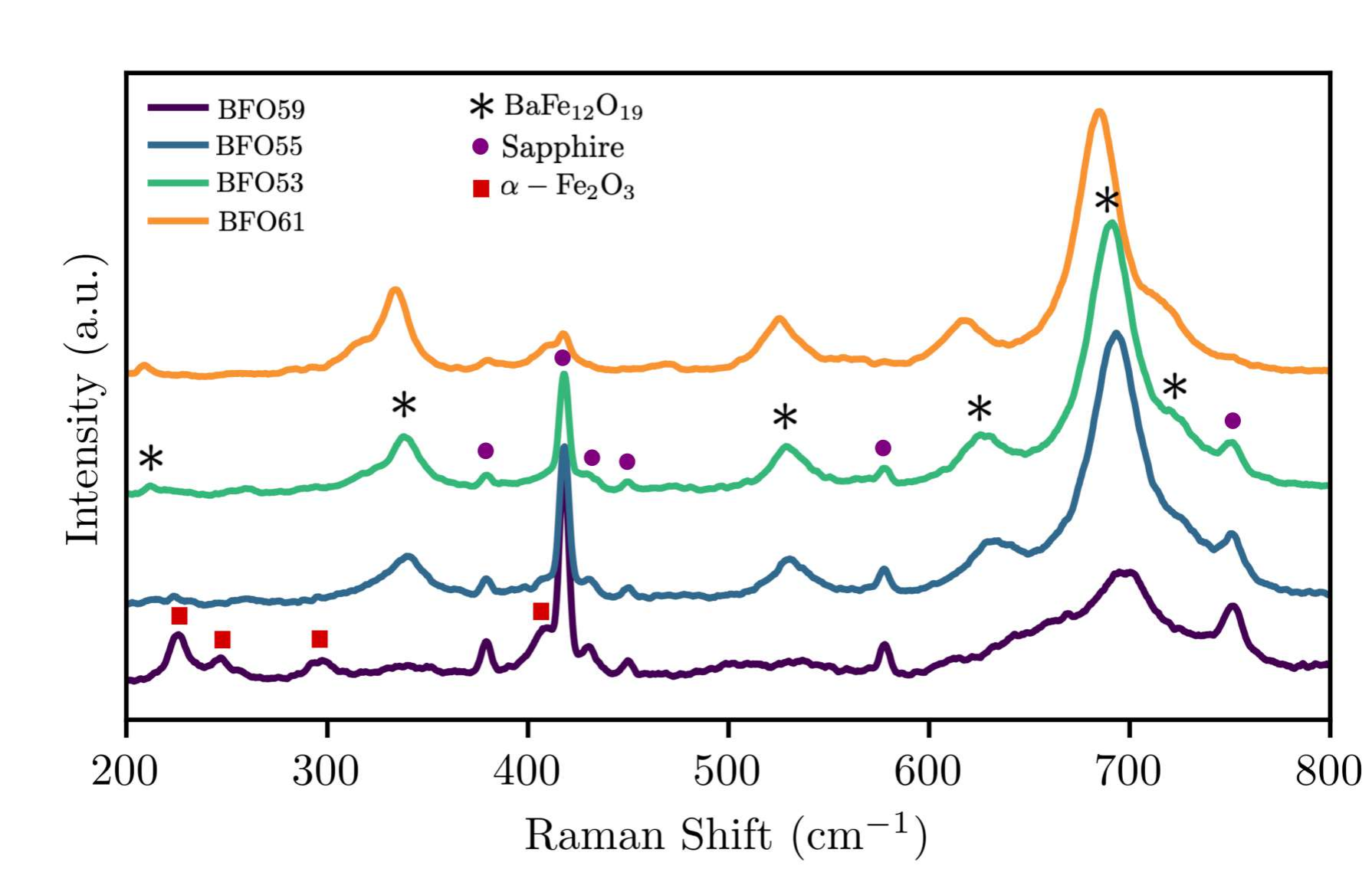}
\caption{Raman spectra of the BaFe$_{12}$O$_{19}$ thin films, showing the characteristic Raman-active modes of the different samples.}
\label{fig:S4}
\end{figure}

Samples BFO55 and BFO53 exhibit the characteristic Raman modes of magnetoplumbite BaFe$_{12}$O$_{19}$, with prominent $E_{2g}$ ($\sim$340 cm$^{-1}$), $E_{1g}$ ($\sim$530 cm$^{-1}$), and $A_{1g}$ ($\sim$630, 691, and 720 cm$^{-1}$) modes, indicating the formation of phase-pure BaFe$_{12}$O$_{19}$ without detectable secondary phases. For the thickest film, BFO61, a systematic redshift of the Raman modes is observed, consistent with the strain relaxation inferred from the increase in the out-of-plane lattice parameter. Moreover, the increased film thickness suppresses several Raman features originating from the Al$_2$O$_3$ substrate. In contrast, BFO59 exhibits pronounced Raman modes associated with $\alpha$-Fe$_2$O$_3$, including the $A_{1g}$ mode at $\sim$225 cm$^{-1}$ and the $E_g$ modes at $\sim$245, 290, and 412 cm$^{-1}$, providing clear evidence for the presence of a secondary hematite phase. These observations are in excellent agreement with the x-ray diffraction results.

\section{Magnetization measurements of different samples}

\begin{figure}[h!]
\includegraphics[width=\columnwidth]{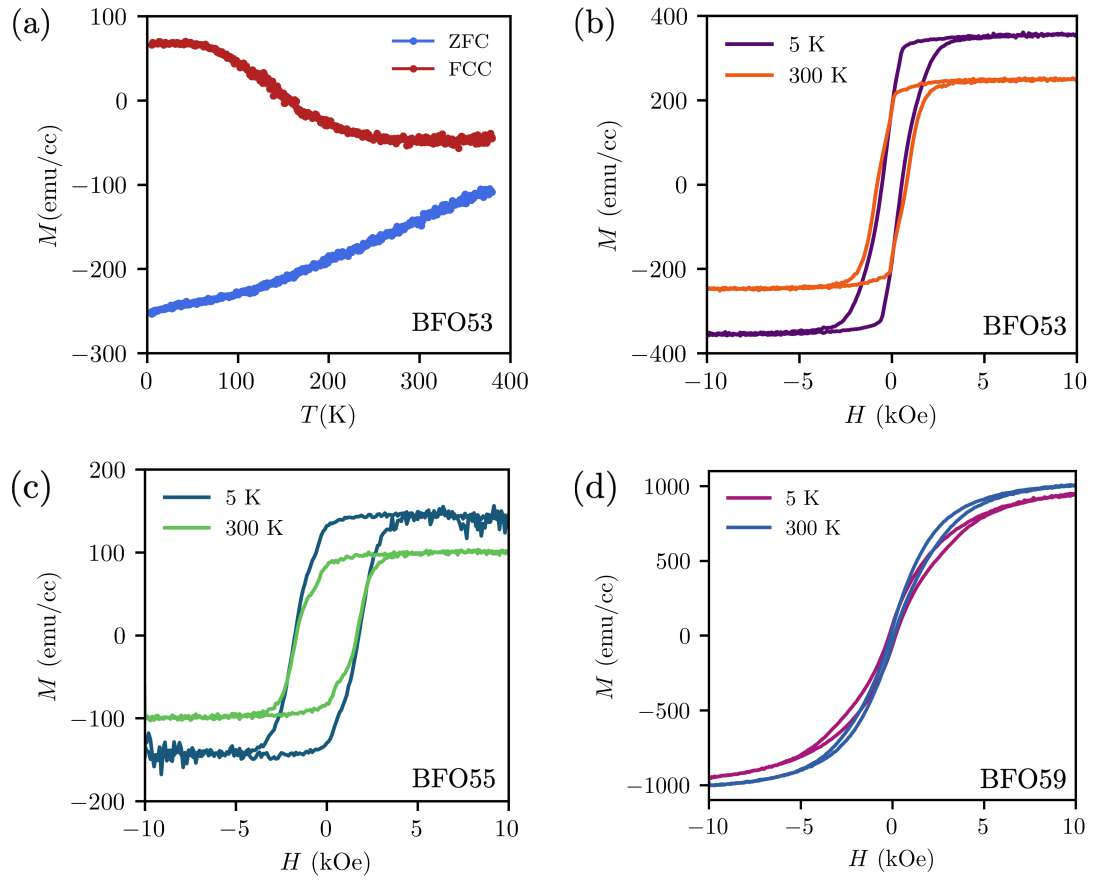}
\caption{ (a) Zero-field-cooled (ZFC) and field-cooled cooling (FCC) magnetization curves of the BFO53 thin film measured with the magnetic field applied along the out-of-plane direction. Out-of-plane magnetic hysteresis loops measured at 5 and 300 K for (b) BFO53, (c) BFO55, and (d) BFO59.}
\label{fig:S5}
\end{figure}

The magnetic measurements reveal a strong dependence of the perpendicular magnetic anisotropy (PMA) on the growth conditions. In particular, BFO59 exhibits a substantially reduced coercivity, indicative of PMA degradation, consistent with the reduced phase purity inferred from the structural and Raman analyses. In contrast, samples BFO53 and BFO55 display well-defined square hysteresis loops characteristic of a pronounced out-of-plane easy axis. The out-of-plane temperature dependence of magnetizations with zero field cooling (ZFC) and field cooled cooling (FCC) under an applied magnetic field of 5000 Oe in the temperature range of 2 - 380 K exhibits ferromagnetic (FM) dominance, indicating a Curie temperature (T$_C$) exceeding 400 K. This observation corroborates the ferrimagnetic characteristics of BaM thin films.

For BFO53, the coercive field increases from $H_c=531$ Oe at 5 K to $H_c=791$ Oe at 300 K, accompanied by an increase in the squareness ratio from 53.7\% to 74.3\%. Interestingly, the enhancement of both coercivity and squareness with increasing temperature suggests a temperature-dependent evolution of the magnetization reversal process and magnetic anisotropy. Furthermore, the in-plane magnetic hysteresis measurements presented in Fig.~2(c) of the main text clearly demonstrate the existence of a perpendicular magnetic easy axis in BFO53. On the other hand, BFO55 exhibits significantly larger coercive fields of 1756.2 and 1659 Oe at 5 and 300 K, respectively, together with high squareness ratios of 93.6\% and 86.9\%. The nearly rectangular hysteresis loops indicate a strong PMA and efficient magnetization reversal along the easy axis.

\section{Details of the BaM/Pt Heterostructure}

\begin{figure}[!ht]
\centering
\includegraphics[width=\columnwidth]{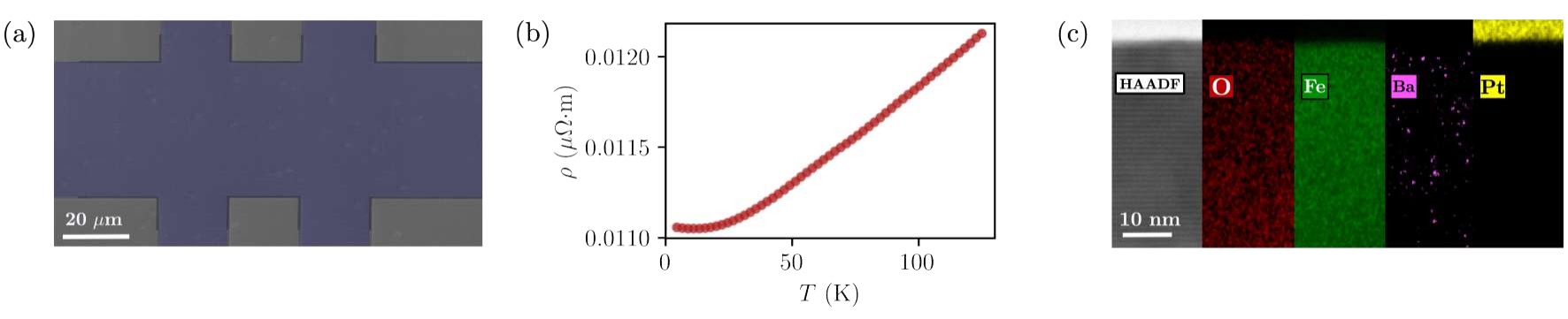}
\caption{(a) Scanning electron microscopy (SEM) image of the BaM/Pt heterostructure. (b) Temperature dependence of the longitudinal resistance of the heterostructure. (c) Cross-sectional energy-dispersive X-ray spectroscopy (EDS) elemental map of the the heterostructure.}
\label{fig:S6}
\end{figure}

The BaM thin film used for the BaM/Pt heterostructure was grown under the same deposition conditions as sample BFO53, with the corresponding growth parameters summarized in Table\ref{tab:deposition}. Subsequently, a 5 nm thick Pt layer was deposited on the BaM thin film at room temperature using an electron-beam evaporator under a base pressure of $3 \times 10^{-6}$ mbar. The Hall bar device was subsequently patterned by focused ion beam (FIB) milling with channel dimensions of $100~\mu\mathrm{m} \times 40~\mu\mathrm{m}$. Fig.\ref{fig:S6}(a) shows the SEM image of the fabricated Hall bar, where the colored region highlights the area patterned by FIB milling. The temperature dependence of the resistance is presented in Fig.\ref{fig:S6}(b). The cross-sectional energy-dispersive X-ray spectroscopy (EDS) analysis shown in Fig.\ref{fig:S6}(c) confirms the well-defined BaM/Pt interface with no interdiffusion between the BaM and Pt layers. In particular, no Fe diffusion into the Pt layer is observed.

\section{Details about B\lowercase{a}M/BSTS device}

Bulk $\text{BiSbTeSe}_2$ (BSTS) crystals, purchased from HQ Graphene, were mechanically exfoliated using Scotch tape. The exfoliated flakes were subsequently picked up using PDMS tape and dry-transferred onto the BaM film using a SUSS MJB4 mask aligner. For device patterning, the substrate was spin-coated with a bilayer photoresist consisting of LOR 10A and S1805, followed by optical lithography. A Ti/Au metal contact stack was deposited via electron-beam evaporation, and metal lift-off was performed in TechniStrip MLO at 75 °C. The devices were packaged inside a chip carrier, wire-bonded, and subsequently measured in a variable temperature insert (VTI). Flake thickness was determined using atomic force microscopy (AFM, Bruker).

\begin{figure}[ht]
\centering
\includegraphics[width=\columnwidth]{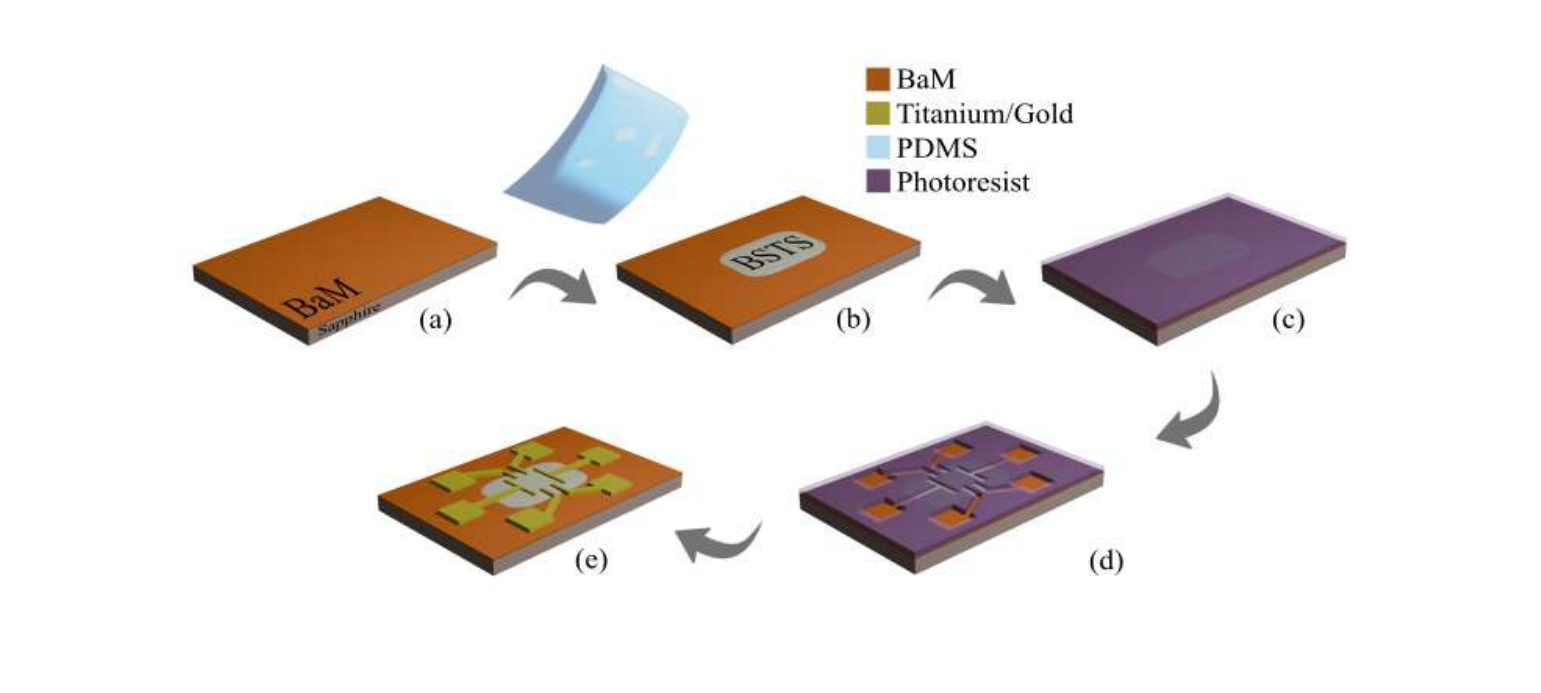}
\caption{Fabrication process flow for the BaM/BSTS device: (a) Growth of BaM on a sapphire (0001) substrate; (b) Dry-transfer of exfoliated flakes from the PDMS tape onto the BaM substrate; (c) Photoresist spin-coating; (d) Lithographic patterning; (e) Final device after Ti/Au metal deposition and lift-off.}
\label{fig:S7}
\end{figure}

\end{document}